\documentclass{aa}

\usepackage{graphicx}
\usepackage{txfonts}
\usepackage{natbib}
\usepackage{amsmath} 
\usepackage{amssymb}
\usepackage{mathtools}
\usepackage{microtype}
\usepackage[dvipsnames]{xcolor}
\definecolor{CiteRed}{RGB}{110, 0, 0}
\usepackage{multirow}
\usepackage{array}
\usepackage{makecell}
\usepackage{siunitx}
\usepackage{xcolor}
\usepackage{orcidlink}
\usepackage{graphicx}
\usepackage{subcaption}
\usepackage{verbatim}
\defcitealias{SternSvensson96}{SS96}
\defcitealias{Kobayashi97}{K97}
\defcitealias{Guidorzi15b}{G15}
\defcitealias{Ghirlanda22}{GS22}
\defcitealias{Bazzanini24}{B24}
\defcitealias{Maistrello25}{M25}

\usepackage{hyperref}
\hypersetup{
    breaklinks,
    colorlinks,
    citecolor=CiteRed,
    linkcolor=NavyBlue,
    urlcolor=NavyBlue,
    linktoc=page,
    pdftitle={},
    pdfkeywords={GRBs, Machine Learning, Genetic Algorithms},
    pdfauthor={Manuele Maistrello},
    pdfcreator={\LaTeX}
}
\usepackage{placeins}

\begin{document} 

   \title{An internal shock model calibrated with real gamma-ray burst light curves using a genetic algorithm}
   
   \author{
        M.~Maistrello~\inst{\ref{unife},\ref{oabo}}\fnmsep\thanks{\texttt{mstmnl[at]unife[dot]it}}\orcidlink{0009-0000-4422-4151}\and
        C.~Guidorzi~\inst{\ref{unife},\ref{oabo},\ref{infnfe}}\orcidlink{0000-0001-6869-0835}\and        S.~Kobayashi\inst{\ref{ljmu}}\orcidlink{0000-0001-7946-4200}\and
        R.~Maccary~\inst{\ref{unife},\ref{oabo}}\orcidlink{0000-0002-8799-2510}
    }
          
   \institute{
        Department of Physics and Earth Science, University of Ferrara, via Saragat 1, I--44122, Ferrara, Italy\label{unife}\and 
        INAF -- Osservatorio di Astrofisica e Scienza dello Spazio di Bologna, Via Piero Gobetti 101, I-40129 Bologna, Italy\label{oabo}\and 
        INFN -- Sezione di Ferrara, via Saragat 1, I--44122, Ferrara, Italy\label{infnfe}\and
        Astrophysics Research Institute, LJMU, IC2, Liverpool Science Park, 146 Brownlow Hill, Liverpool L3 5RF, UK\label{ljmu}
   }

  \date{Received xxx; accepted xxx}

  \abstract
   {The origin of prompt emission in gamma-ray bursts (GRBs) remains a fundamental open question. The internal shock (IS) model is a leading mechanism proposed to explain the dissipation of relativistic ejecta kinetic energy into gamma-rays. However, the model parameters have yet to be fully optimised to reproduce the diverse morphological properties observed in real GRB light curves (LCs).}
   {Utilising a machine-learning framework, we evaluate the IS model through parameter optimisation, comparing the statistical properties of simulated LCs against those of observed data.}
   {Our dataset consists of three GRB catalogues (\textit{Swift}/BAT, \textit{Fermi}/GBM, \textit{CGRO}/BATSE). By adopting a model for the GRB formation rate as a function of redshift, we employed a genetic algorithm to optimise the IS model parameters. The algorithm minimises a total loss function based on six independent metrics, representing both the average properties and the statistical distributions of real LCs.}
   {The calibrated IS model successfully reproduces the average post-peak GRB temporal profile, together with the corresponding root-mean-square and third-moment temporal profiles, as well as the average autocorrelation function. Furthermore, it recovers the observed distributions of duration, signal-to-noise ratio, peak count per burst, peak flux, and fluence. We find that a generalised Zipf distribution governs the number of shells emitted per GRB, while rest-frame emission times follow a negative exponential distribution.}
   {The optimised formulation of the IS model reproduces a wide range of observed GRB LC properties, despite its simplified treatment of radiation physics. Moreover, it provides key insights into the central engine's activity: (i) the emission times suggest a stochastic process where all shells within a given burst have an identical, independent, constant probability of ejection per unit time; (ii) the heavy-tailed distribution of the number of shells per GRB mirrors the frequency-magnitude distribution of earthquakes, known as the Gutenberg-Richter law. Finally, this optimised model serves as a predictive tool for the GRB populations expected to be detected by future missions.}

   \maketitle

\section{Introduction}
\label{sec:introduction}

Gamma-ray bursts (GRBs) are intense flashes of gamma rays, ranging in duration from milliseconds to thousands of seconds, originating at cosmological distances. They are linked to catastrophic stellar events, including the collapse of massive, hydrogen-stripped stars---known as `collapsars' \citep{Woosley93, Paczynski98, MacFadyen99}---and the merger of compact binaries \citep{Paczynski86, Eichler89, Abbott17}. In both scenarios, the progenitor leaves behind a central engine---either a hyper-accreting stellar-mass black hole (BH; \citealp{Popham99}) or a rapidly rotating, highly magnetised neutron star (NS; \citealp{Usov92})---, which launches an ultra-relativistic jet. 
While the GRB emission is believed to arise from internal dissipation within this outflow, several fundamental questions remain. The nature of the central engine, the jet-launching mechanism, the jet composition, and the specific processes that convert internal energy into gamma rays are still debated. This uncertainty motivates the development of physically grounded models capable of reproducing the diverse temporal variability observed in GRB light curves.

One of the earliest models proposed to explain the GRB emission---commonly referred to as the prompt emission---is the so-called `internal shock' (IS) model \citep{Rees94}. Within this framework, an initially hot fireball composed of photons, electron-positron pairs, and a small amount of baryons converts most of its thermal energy into kinetic energy, which is subsequently partially dissipated through mildly relativistic shocks. These shocks are thought to develop within a highly variable relativistic outflow driven by an intermittent central engine. Electrons are accelerated at the shock fronts, and the resulting internal energy is radiated via synchrotron and inverse Compton processes, powering the observed prompt emission. 
The identification of a thermal component superimposed on the non-thermal spectrum in a few GRBs has spurred the development of photospheric models. In these models, a fraction of the prompt emission energy originates at the photosphere, where the jet becomes optically thin, resulting in emission at significantly smaller radii than predicted by the standard IS model (see \citealt{Peer15rev, BeloborodovMeszaros17} for reviews).

In contrast to alternative scenarios, such as the ICMART (internal-collision-induced magnetic reconnection and turbulence; \citealp{ICMART}) model, the magnetic field in the IS framework is not assumed to play a dynamically dominant role in the ejecta, thereby providing a physically motivated framework to link the variability of the central engine to the observed temporal structure of GRB light curves (LCs). Previous studies have indeed shown a strong correlation between the occurrence time of a pulse in a GRB LC and the ejection time of the corresponding shell from the central engine. The number of pulses in a GRB LC reflects the number of shells ejected by the engine, while the separation between pulses traces the time intervals during which the engine was inactive. Thus, the observed temporal structure directly encodes the activity of the central engine  (\citealp{Kobayashi97}, hereafter \citetalias{Kobayashi97}). 

In our previous studies (\citealp{Bazzanini24, Maistrello25}; hereafter \citetalias{Maistrello25}), we presented a well-calibrated workflow to optimise the parameters of a stochastic pulse-avalanche model for GRB LCs \citep{SternSvensson96}, starting from real data. The workflow employs a machine-learning technique known as genetic algorithm (GA), enabling the generation of synthetic LCs that successfully reproduce a set of average temporal properties of real GRB LCs observed by the Burst Alert Telescope (BAT; \citealt{Barthelmy05}) aboard the Neil Gehrels Swift Observatory, the Gamma-ray Burst Monitor (GBM; \citealt{Meegan09}) aboard \textit{Fermi}, and the Burst and Transient Source Experiment (BATSE; \citealt{Paciesas99}) aboard the Compton Gamma Ray Observatory (CGRO; 1991--2000). However, the stochastic pulse-avalanche model lacks a direct physical connection to the underlying dissipation mechanism. A key open question is therefore whether a physically motivated model, such as the IS scenario, can reproduce the observed temporal properties when directly calibrated against real data.

To our knowledge, this is the first systematic optimisation of an IS model directly against multiple statistical properties of observed GRB LCs. To this end, we implement a version of the IS model of \citetalias{Kobayashi97}, optimised following the prescription described in \citetalias{Maistrello25}. The paper is organised as follows. In Sect.~\ref{sec:data_analysis}, we describe the sample selection and the metrics adopted for the model optimisation. The model is presented in Sect.~\ref{sec:model}. Results, discussion, and conclusions are reported in Sections~\ref{sec:results}, ~\ref{sec:discussion}, and \ref{sec:conc}, respectively.

\section{Data analysis}
\label{sec:data_analysis}

\subsection{Sample selection}
\label{ss:sample_selection}

We considered three complementary GRB datasets, drawn from as many instruments and energy passbands: (i) 2024 GRBs from the 4B BATSE catalogue \citep{Paciesas99}; (ii) 1389 GRBs detected by \textit{Swift}/BAT between January 2005 and November 2023; (iii) 2356 GRBs from the fourth \textit{Fermi}/GBM catalogue \citep{vonKienlin20}. Hereafter, these datasets are referred to as the BATSE, \textit{Swift}, and \textit{Fermi} samples, respectively.

For all samples, we analysed background-subtracted LCs in the total instrumental passbands, namely 25--2000 keV for BATSE, 15--150 keV for \textit{Swift}/BAT, and 8--1000 keV for \textit{Fermi}/GBM, adopting a uniform time resolution of 64 ms. For the \textit{Fermi} sample, we used the full NaI scintillator passband and performed background subtraction with the publicly available GBM Tools\footnote{\url{https://fermi.gsfc.nasa.gov/ssc/data/analysis/gbm/gbm_data_tools/gdt-docs/}}, following the procedure described in \citet{Maccary24b}.

Only long-duration GRBs were retained by imposing a threshold of $T_{90} > 2$ s. Additional cuts were applied on the signal-to-noise ratio (S/N), requiring S/N $\ge 15$, $\ge 10$, and $\ge 15$ for the BATSE, \textit{Swift}, and \textit{Fermi} samples, respectively. Consequently, the samples (i)--(iii) were reduced to 1389, 635, and 1120 GRBs, respectively. The adopted thresholds were chosen to preserve the diversity of LC morphologies while ensuring adequate statistical quality. 

Events showing compelling evidence for a compact-object merger origin were excluded from each dataset: GRB\,060614 \citep{DellaValle06,Jin15}; GRB\,211211A \citep{Rastinejad22,Troja22,Yang22}; GRB\,191019A \citep{Levan23,Stratta25}. This choice preserves the homogeneity of the sample by excluding events with a different physical origin from the bulk of the population, although the number of long-duration GRBs associated with compact-object mergers may be larger than what common wisdom assumed thus far \citep{Maccary26b}.

\subsection{Statistical metrics}
\label{ss:metrics}

To quantify the agreement between simulated and real LCs, we adopted the set of metrics introduced in \citetalias{Maistrello25}. These metrics capture complementary aspects of GRB LCs and are briefly summarised below (see \citetalias{Maistrello25} and references therein for a detailed discussion):
\begin{itemize}
    \item The average post-peak time profile, $\langle F/F_{\rm p}\rangle$, obtained by aligning LCs at their brightest peak and averaging over the first 150 s, where $F_{\rm p}$ denotes the peak count rate.
    \item The average third moment of the post-peak profile, $\langle(F/F_{\rm p})^3\rangle$, which is sensitive to profile asymmetries.
    \item The average autocorrelation function (ACF) of the LCs.
    \item The distribution of LC durations expressed in terms of $T_{20\%}$, defined as the time interval during which the signal exceeds 20\% of its maximum value \citep{SternSvensson96}.
    \item The distribution of the S/N of the LCs.
\end{itemize}

Additionally, we included the distribution of the number of peaks per GRB, which showed evidence for the existence of two regimes in long GRBs---namely peak-poor and peak-rich behaviours---consistently in all three datasets \citep{Guidorzi24, Maccary24b}. It complements the original set by constraining the temporal complexity of the simulated GRBs.

\section{Internal shock model}
\label{sec:model}

The LCs were generated using an IS model originally introduced by \citetalias{Kobayashi97}. In the following, we outline the basic elements of the model that are relevant for our simulations.

ISs develop within a relativistic outflow made up of multiple shells ejected with different Lorentz factors. Energy dissipation occurs when a faster shell catches up with a slower one, leading to an inelastic collision and the formation of a merged shell. Consistently with the notation adopted by \citetalias{Kobayashi97}, physical quantities associated with the rapid shell, the slow shell, and the merged shell are denoted by the subscripts $r$, $s$, and $m$, respectively.

Under the assumption of a completely inelastic collision, the Lorentz factor of the merged shell is given by
\begin{equation}
    \gamma_m \simeq \sqrt{\dfrac{m_r\gamma_r + m_s\gamma_s}{m_r/\gamma_r + m_s/\gamma_s}}\,,
\end{equation}
where $m_r$ and $m_s$ are the shell masses, and $\gamma_r\gg 1$ and $\gamma_s\gg 1$ are the corresponding Lorentz factors. The internal energy generated during the collision is obtained from the difference between the kinetic energy of the system before and after the interaction,
\begin{equation}
    E_{\mathrm{int}} = m_rc^2(\gamma_r - \gamma_m) + m_sc^2(\gamma_s - \gamma_m)\,.
\end{equation}
Assuming that this internal energy is promptly radiated isotropically in the comoving frame of the merged shell, the efficiency of conversion from kinetic to internal energy can be expressed as
\begin{equation}
    \epsilon = 1 - \dfrac{(m_r + m_s)\gamma_m}{m_r\gamma_r + m_s\gamma_s}\,.
\end{equation}

The observed emission from a single shell collision appears as a pulse whose temporal width $\delta T$ is determined by several characteristic timescales, namely the radiative cooling time, the hydrodynamic time, and the angular spreading time. Since the internal energy is released via synchrotron and inverse Compton processes, and electron cooling occurs on timescales much shorter than the hydrodynamic one, radiative cooling can be neglected in this context.
The hydrodynamic timescale corresponds to the time required for the shocks generated during the collision to cross the shells. The interaction produces both a forward and a reverse shock, and the emission duration in the comoving frame is approximately the time needed for the reverse shock to cross the rapid shell,
\begin{equation}
    \delta t_e = \dfrac{l_r}{c(\beta_r - \beta_{rs})}\,,
\end{equation}
where $l_r$ is the width of the rapid shell, while $\beta_r$ and $\beta_{rs}$ denote the velocities of the rapid shell and of the reverse shock, respectively. Due to relativistic motion of the emitting region towards the observer, the corresponding emission timescale in the observer frame is reduced by a factor $1/2\gamma_m^2$.

If the collision is assumed to occur at $t=0$ at a sufficiently large radius $R$, angular spreading effects play a role in shaping the observed pulse. In particular, when the initial separation $L$ between the shells exceeds $l_r$, angular effects broaden the pulse and produce an asymmetric temporal profile characterised by a fast rise and a slower decay. In this case, the observed luminosity can be written as
\begin{equation}
    \label{eq:pulse}
    {\cal L} =
    \begin{cases}
        0 ~~~~~~~~~~ (t \leq 0)\,, \\
        h\left[1 - \dfrac{1}{(1 + 2\gamma_m^2 ct/R)^2}\right] ~~~~~~~~~~\Big(0 < t \leq \dfrac{\delta t_e}{2\gamma_m^2}\Big)\,, \\
        h\left\{\dfrac{1}{[1 + (2\gamma_m^2t - \delta t_e)c/R]^2} - \dfrac{1}{(1 + 2\gamma_m^2 ct/R)^2} \right\} ~~ \Big(t > \dfrac{\delta t_e}{2\gamma_m^2}\Big)\,,
    \end{cases}
\end{equation}
where $h = E_{\mathrm{int}}2\gamma_m^2/\delta t_e$. Following the shock crossing, the two initial shells merge into a thinner shell of width
\begin{equation}
    l_m = l_s\dfrac{\beta_{fs} - \beta_m}{\beta_{fs} - \beta_s} + l_r\dfrac{\beta_m - \beta_{rs}}{\beta_r - \beta_{rs}}\,,
\end{equation}
with $\beta_s$ and $\beta_{fs}$ representing the velocities of the slower shell and of the forward shock, respectively. Both $\beta_{fs}$ and $\beta_{rs}$ are calculated following \citet{Sari95} and \citetalias{Kobayashi97}.

In a realistic outflow, a large number of shell collisions occur. Each interaction produces a shock that converts a fraction of the kinetic energy into thermal energy, which is subsequently radiated through synchrotron and inverse Compton emission, giving rise to a pulse described by Eq.~\eqref{eq:pulse}. The observed LC is then obtained as the superposition of the pulses produced by all individual collisions.

The outflow is modelled as a wind of $n$ shells, where $n$ is sampled from a generalised Zipf distribution $p(n)$\footnote{Also known as Zipf-Mandelbrot distribution.} \citep{Mandelbrot1953}:%
\begin{equation}
    p(n)\ =\ \frac{K}{(n + S)^\alpha}\;,~~~~(2\le n\le n_{\rm max})\;,
\label{eq:zipf}
\end{equation}
where $K$ is a normalisation constant, $n_{\rm max}$ is the maximum value (in our model, it is set to a few hundreds), and two free parameters, $S$ (called `shift') and $\alpha$. This law is found to describe many ranking distributions in the context of complex systems \citep{DeMarzo21}; in physics, for instance, it describes the distribution of the size of earthquakes of a given geographical region as a function of their ranking, which is sorted according to the magnitude, the so-called Gutenberg-Richter law \citep{Gutenberg44}.
The choice of Eq.~\eqref{eq:zipf} followed a number of alternative and less successful attempts: in particular, a heavy-tailed distribution turned out to be a key requirement.

Once, for a given GRB, $n$ is drawn from Eq.~\eqref{eq:zipf}, the corresponding emission times are drawn from a common exponential probability density function (PDF):
\begin{equation}
    f(t) = \frac{1}{\tau}\ \exp\left(-\dfrac{t}{\tau}\right)\,,
\label{eq:exp_t}
\end{equation}
where $\tau$ is different for each GRB and results from sampling a log-normal distribution: $\log{\tau}\sim {\cal N}(\mu_\tau, \sigma^2_\tau)$, where $\mu_\tau$ and $\sigma_\tau$ are treated as free parameters by the model. Equation~\eqref{eq:exp_t} implies a temporally declining shell emission rate, which differs from the constant one assumed by \citetalias{Kobayashi97}. This is one of the two major changes adopted in the present model with respect to the original formulation of \citetalias{Kobayashi97}. The rationale behind Eq.~\eqref{eq:exp_t} follows the simple idea that each shell has the same independent and constant probability per unit time of being emitted: the resulting distribution is precisely Eq.~\eqref{eq:exp_t}, similarly to the nuclear decay of a given number of atoms of a radioactive element. This simple process was first proposed by \citet{Guidorzi15b} to explain the observed waiting time distribution of gamma-ray peaks and GRB early X-ray flares. The property of a declining shell emission rate agrees with the picture of a central engine that behaves like an initially out-of-equilibrium system and progressively evolves towards a more stable configuration. 

Each shell is characterised by three quantities: the Lorentz factor $\gamma_i$, the mass $m_i$, and the width $l_i$, with $i = 1, \ldots, n$. As shown by \citetalias{Kobayashi97}, the efficiency of the model in converting kinetic energy into internal energy increases with the ratio $\gamma_{\max}/\gamma_{\min}$. The second important change with respect to the \citetalias{Kobayashi97} model concerns the way the Lorentz factors of the shells for the $j$th GRB are sampled: we first draw two values, $\gamma_{\min}^{(j)}$ and $\gamma_{\max}^{(j)}$, within the interval $[\gamma_{\min}, \gamma_{\max}]$, and then assume that $\gamma_i$ is distributed uniformly in logarithmic space between $\gamma_{\min}^{(j)}$ and $\gamma_{\max}^{(j)}$. This prescription introduces burst-to-burst variability in the Lorentz factor contrast while preserving a broad overall range set by $\gamma_{\min}$ and $\gamma_{\max}$.

The shell mass is assumed to depend on the Lorentz factor according to $m_i = p\gamma_i^{\eta}$, with $\eta = -1$ and $p$ a proportionality constant. This choice corresponds to shells carrying equal energy, such that lighter shells are faster. Finally, to reduce the dimensionality of the parameter space, we assume a common ratio $r = c\Delta t / l$, where $\Delta t$ is the time interval between two consecutive shells and $l$ the width of the later shell, effectively linking the shell separation to their width. 

Under these assumptions, the free parameters of the model are $\alpha$, $S$, $\mu_\tau$, $\sigma_\tau$, $\gamma_{\min}$, $\gamma_{\max}$, $\log p$, and $\log r$. The parameters are summarised in Table~\ref{tab:parameters}, together with their exploration ranges and corresponding priors.

The dynamical evolution of the system is implemented following \citetalias{Kobayashi97}; additional details are provided in Appendix~\ref{app:multiple_shell_model}. The final temporal profile is obtained as the superposition of the pulses produced by the individual two-shell interactions described by Eq.~\eqref{eq:pulse}. This represents the theoretical LC expressed in terms of bolometric luminosity, as measured by an observer at rest in the GRB comoving frame.

A redshift value $z_{\mathrm{GRB}}$ is then assigned to each simulated event from sampling the GRB rate $R(z)$:
\begin{equation}
    R(z)\ = \frac{R_{\rm grb}(z)}{1+z}\,\frac{dV(z)}{dz}
    \label{eq:Rz}
\end{equation}
where $R_{\rm grb}(z)$ is the GRB formation rate taken from \citet[hereafter GS22]{Ghirlanda22},
\begin{equation}
    R_{\rm grb}(z)\ = R_{\rm grb}(0)\ \frac{(1 + z)^{p_{z,1}}}{1 + \Big(\frac{1+z}{p_{z,2}} \Big)^{p_{z,3}}}
    \label{eq:Rgrbz}
\end{equation}
where $R_{\rm grb}(0)$ is the local volumetric rate and $p_{z,1}=3.33$, $p_{z,2}=3.42$, and $p_{z,3}=6.21$. The comoving volume element is given by
\begin{equation}
    \dfrac{dV(z)}{dz} = \dfrac{4\pi\,c\,d_L^2(z)}{H(z)\,(1 + z)^2}\,,
    \label{eq:dV_dz}
\end{equation}
where $d_L(z)$ is the luminosity distance and $H(z) = H_0[\Omega_M(1 + z)^3 + \Omega_\Lambda + (1 - \Omega_M - \Omega_\Lambda)(1 + z)^2]^{1/2}$. The $(1 + z)$ term in the denominator of Eq.~\eqref{eq:Rz} accounts for the cosmic dilation in the observed rate. The redshift is randomly drawn from Eq.~\eqref{eq:Rz}, once it is properly normalised as a probability density function.

The assigned redshift $z_{\mathrm{GRB}}$ is used to compute the luminosity distance and convert the bolometric luminosity LC into an observed flux LC within the instrumental energy passband. The rest-frame photon spectrum $N(E)$ is assumed to follow a Band function \citep{Band93}, characterised by the intrinsic peak energy $E_{\mathrm{p,i}}$ and the low- and high-energy spectral indices $\alpha$ and $\beta$. The value of $E_{\mathrm{p,i}}$ is derived from the $E_{\mathrm{p,i}}$--$L_{\gamma,\rm{iso}}$ correlation reported in \citetalias{Ghirlanda22}.  The spectral indices are randomly drawn from the corresponding observed distributions (see Appendix~\ref{app:spectral_parameters} for details). The $k$-correction factor is then computed as
\begin{equation}
    k_{\mathrm{flux}} = \dfrac{\int_{E_1(1 + z_{\mathrm{GRB}})}^{E_2(1 + z_{\mathrm{GRB}})}EN(E)dE}{\int_{1}^{10^4}EN(E)dE}\,,
\end{equation}
where $E_1$ and $E_2$ define the limits of the instrumental energy passband. This expression holds under the assumption of no spectral evolution.

The flux LC, $F$, is subsequently converted into counts (integrated over 64 ms) or count rates, $R$, depending on the instrument, following the methodology described in \citetalias{Maistrello25}. For completeness, we briefly summarise the procedure here. The conversion is expressed through a factor $k = \log_{10}(F/R)$, where $F$ is the flux and $R$ represents counts for BATSE, count rate per fully illuminated detector for an equivalent on-axis source in the case of \textit{Swift}/BAT, and counts per detector for \textit{Fermi}/GBM. The conversion factors were derived using the fluence values reported in the official GRB catalogues: the fourth BATSE catalogue \citep{Paciesas99}, the third \textit{Swift}/BAT catalogue \citep{Lien16}, and the fourth \textit{Fermi}/GBM catalogue \citep{vonKienlin20}. The corresponding counts were obtained by integrating the signal over the 7$\sigma$ time interval, defined as the interval between the first and last bins in which the signal exceeds the background by at least $7\sigma$.

As a final step, background contributions are added to the simulated LCs. Following \citetalias{Maistrello25}, constant background rates of 2.9 counts s$^{-1}$ and 39.4 counts s$^{-1}$ were adopted for BATSE and GBM, respectively. These values correspond to the median measured error rates of the respective real LCs. The final LCs are generated as Poisson realisations with expectation value equal to the sum of the noise-free LC and the background level; the background is subsequently subtracted. For \textit{Swift}/BAT, each bin is instead sampled from a Gaussian distribution with mean equal to the noise-free LC and standard deviation extracted from the distribution of the real LC uncertainties.

The optimisation of the model parameters was carried out using a GA, following the same general framework adopted in \citetalias{Maistrello25} (see Appendix~\ref{app:ga} for details). 

\section{Results}
\label{sec:results}

\begin{table}
    \small
    \setlength{\tabcolsep}{3pt}
    \renewcommand{\arraystretch}{1.4}
    \caption{Best-fitting parameters of the IS model for the \textit{Swift}/BAT, \textit{Fermi}/GBM, and \textit{CGRO}/BATSE samples.}
    \label{tab:optimised_parameters}
    \centering 
        \begin{tabular}{c | c c c}
        \hline\hline
        Parameter                                      & \textit{Swift}/BAT     & \textit{Fermi}/GBM    & \textit{CGRO}/BATSE\\
                                                       & (15--150 keV)          & (8--1000 keV)         & (25--2000 keV)\\
        \hline
        $\alpha$                                       & $2.06^{+0.01}_{-0.01}$ & $2.25^{+0.02}_{-0.05}$ & $2.26^{+0.20}_{-0.06}$\\ 
        $S$                                            & $1.36^{+0.56}_{-0.17}$ & $2.41^{+0.31}_{-0.33}$ & $2.45^{+0.48}_{-0.62}$\\ 
        $\mu_\tau$                                     & $3.53^{+0.01}_{-0.11}$ & $3.14^{+0.14}_{-0.13}$ & $2.76^{+0.10}_{-0.08}$\\ 
        $\sigma_\tau$                                  & $0.41^{+0.36}_{-0.28}$ & $0.25^{+0.19}_{-0.13}$ & $0.30^{+0.31}_{-0.07}$\\ 
        $\gamma_{\min}$                                & $30^{+6}_{-3}$         & $23^{+7}_{-10}$        & $33^{+7}_{-10}$\\ 
        $\gamma_{\max}$                                & $776^{+179}_{-214}$    & $549^{+263}_{-195}$    & $832^{+59}_{-140}$\\ 
        $p\,(10^{32}\,\mathrm{g})$                     & $5.2^{+0.1}_{-0.9}$    & $5.4^{+0.3}_{-1.2}$    & $3.9^{+0.2}_{-0.3}$\\ 
        $r$                                            & $3.1^{+1.1}_{-0.6}$    & $7.2^{+1.9}_{-1.6}$    & $4.6^{+2.5}_{-1.1}$\\
        \hline 
        Loss (Train best)                              & 0.29 & 0.64 & 1.29 \\
        Loss (Train avg.)                              & 0.92 & 1.44 & 2.43 \\\hline 
        Loss (Test)                                    & 0.35 & 1.21 & 2.17 \\\hline 
        Loss (Test: $\langle F / F_{\rm p}\rangle$)    & 0.18 & 0.36 & 0.30 \\
        Loss (Test: $\langle(F / F_{\rm p})^3\rangle$) & 0.12 & 0.13 & 0.27 \\
        Loss (Test: $\langle\mathrm{ACF}\rangle$)      & 0.60 & 1.86 & 0.78 \\
        Loss (Test: $T_{20\%}$)                        & 1.20 & 2.12 & 1.67 \\
        Loss (Test: S/N)                               & 0.00 & 0.00 & 0.00 \\
        Loss (Test: $N_p$)                             & 0.00 & 2.78 & 10.0 \\
        \hline 
        \hline 
        \end{tabular}
        \tablefoot{Parameter values correspond to the medians of the distributions in the final GA generation, with uncertainties given by the 5th and 95th percentiles. `Train best' and `Train avg.' denote the minimum and mean loss values in the final generation, respectively. The test set consists of an independent sample of 5000 simulated LCs. The last rows report the individual contributions of each metric to the total test loss.}
\end{table}

\subsection{Optimal parameter values}
\label{ss:optimal_parameters}

\begin{figure*}[!ht]
   \centering
   \includegraphics[width=1\linewidth]{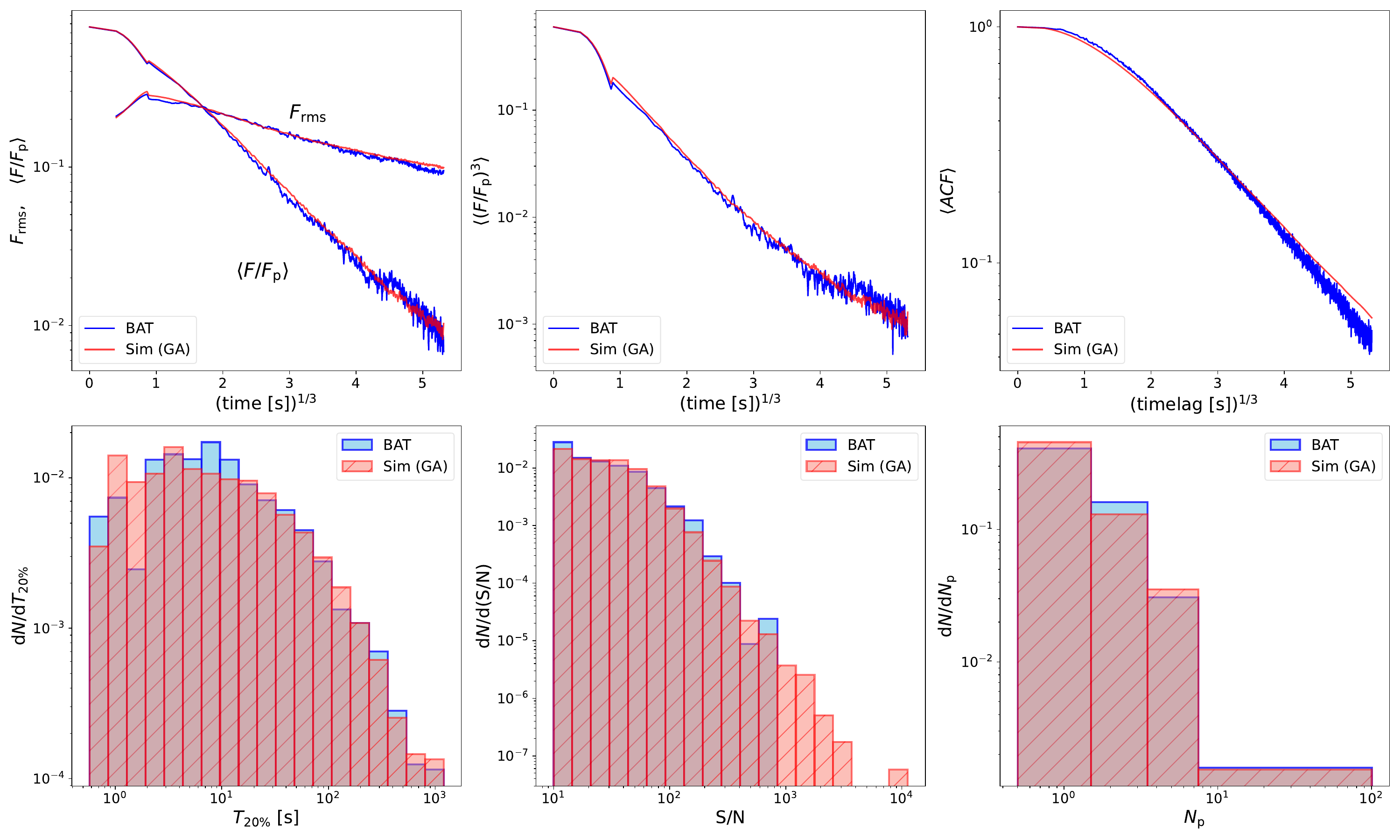}
   \caption{Distributions of the six metrics for real (blue) and simulated (red) LCs of the \textit{Swift}/BAT dataset. The simulated LCs were obtained from the optimised parameter set (Table~\ref{tab:optimised_parameters}).
   \textit{Top left}: average peak-aligned post-peak normalised time profile and corresponding root-mean-square (r.m.s.) deviation of the individual profiles, $F_{\mathrm{rms}} \equiv \big[\langle(F / F_{\rm p})^2\rangle - \langle F / F_{\rm p}\rangle^2\big]^{1/2}$. 
   \textit{Top centre}: average peak-aligned third-moment profile. 
   \textit{Top right}: average autocorrelation function (ACF). 
   \textit{Bottom left}: distribution of the $T_{20\%}$ duration. 
   \textit{Bottom centre}: S/N distribution. 
   \textit{Bottom right}: distribution of the number of peaks per GRB. 
   As in \citetalias{Maistrello25}, the top-row profiles were smoothed using a Savitzky--Golay filter to reduce the effects of Poisson noise.}
   \label{fig:metrics_swift}
\end{figure*}

The optimised model parameters for the \textit{Swift}, \textit{Fermi}, and BATSE samples---defined as the median values of the final-generation population---are listed in Table~\ref{tab:optimised_parameters}. To provide a robust estimate of the model performance, the relative loss values were recomputed using an independent set of 5000 simulated LCs for each dataset. Figure~\ref{fig:metrics_swift} compares the distributions of the six metrics for real and simulated LCs of the \textit{Swift}/BAT sample. Examples of LCs are illustrated in Fig.~\ref{fig:lc_example}. 

Overall, the optimised solutions work perfectly for all three datasets, as indicated by the test loss. Only for BATSE, the model provides a poor result on the sixth metric (distribution of the number of peaks per GRB), while the remaining five metrics are satisfactorily fulfilled. All parameters have very similar values within uncertainties across the different datasets. Notably, $\mu_\tau$ slightly decreases passing from the softest to the hardest energy passband: this is possibly driven by the fact that real LCs tend to last longer in the softer energy channels than in the harder ones.

\subsection{Properties of the simulated GRB populations}
\label{ss:distributions}

We derived the distributions of several fundamental quantities from the optimised IS models and compared them with observations. As far as intrinsic properties are concerned, such as $L_{\gamma,{\rm iso}}$ and $E_{\gamma,{\rm iso}}$, the real distributions are based on the overall sample of GRBs with measured redshift, which is primarily given by \textit{Swift}/BAT, and then \textit{Fermi}/GBM, Konus/\textit{WIND} and \textit{BeppoSAX}. Since the observed sample includes only GRBs with measured redshift, it is subject to selection effects and is not a complete flux-limited sample. The comparisons shown below should therefore be regarded as consistency checks rather than formal fits. When available, $L_{\gamma,{\rm iso}}$ and $E_{\gamma,{\rm iso}}$ values were preferentially taken from the Konus/\textit{WIND} catalogues \citep{KWGRBcat17,KWGRBcat21} and \textit{Fermi}/GBM \citep{vonKienlin20} because of their broad energy band. The comparison between the expected and observed distributions contributes to further test the ability of the optimised IS model to account for the observed properties and, at the same time, to gain clues on the way IS-based inner engines work.

To allow for a direct comparison that includes faint GRBs, we relaxed the S/N threshold and generated a new set of 5000 LCs for each dataset. We then simulated a detection process to determine whether a synthetic GRB would have been detected. Specifically, we applied a well-calibrated peak search algorithm, {\sc fast-mepsa} \citep{Maistrello26}: if at least one peak with S/N $\ge 5$ is identified, the GRB is considered detected. {\sc fast-mepsa} is an optimised version of the {\sc mepsa} algorithm, originally developed to detect peaks in uniformly sampled time series affected by uncorrelated Gaussian noise \citep{Guidorzi15a}. It has proven effective for large-scale analyses, particularly when the detection of faint events is required.
The detection process implemented with {\sc fast-mepsa} emulates a rate trigger logic. As a consequence, it is not sensitive to very long and faint events that could instead be detected through image-trigger techniques, such as those employed by \textit{Swift}/BAT or any other imaging instrument. Nevertheless, the goal is to assess the capability of the model to reproduce the properties of the observed population of GRBs. Developing a detailed image-trigger simulator lies beyond the scope of this work.

\begin{figure}[!ht]
    \resizebox{\hsize}{!}{\includegraphics{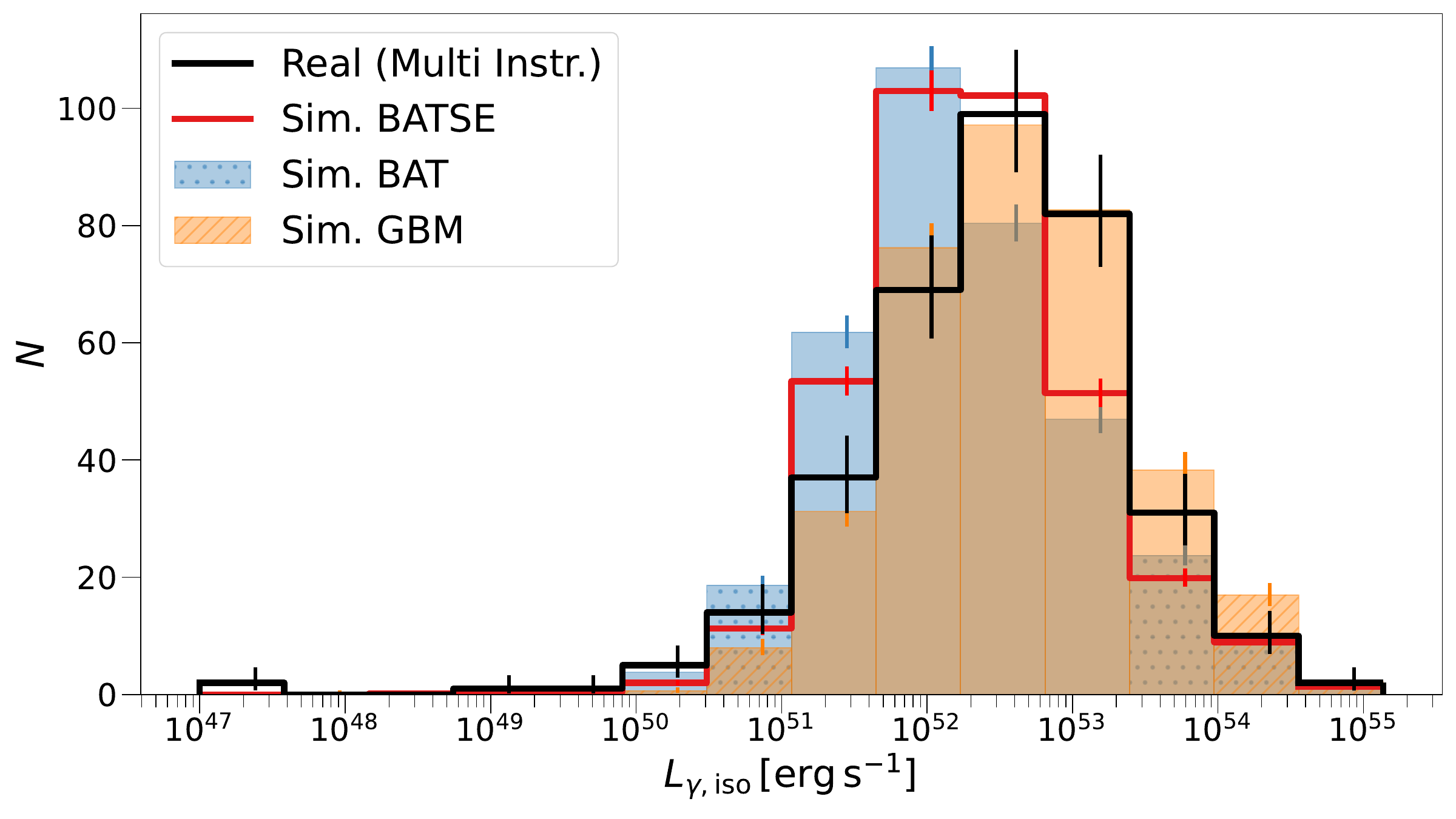}}
    \caption{Bolometric, isotropic-equivalent peak luminosity distributions of the GRB populations simulated for BATSE (red solid line), \textit{Swift}/BAT (shaded blue), and \textit{Fermi}/GBM (shaded orange). The corresponding distribution of real GRBs with known redshift (black solid line) is mostly due to \textit{Swift}, then \textit{Fermi}, Konus/\textit{WIND}, and \textit{BeppoSAX}. The simulated distributions are rescaled to match the total number of observed events.}
    \label{fig:Liso_distribution}
\end{figure}

\begin{figure}[!ht]
    \resizebox{\hsize}{!}{\includegraphics{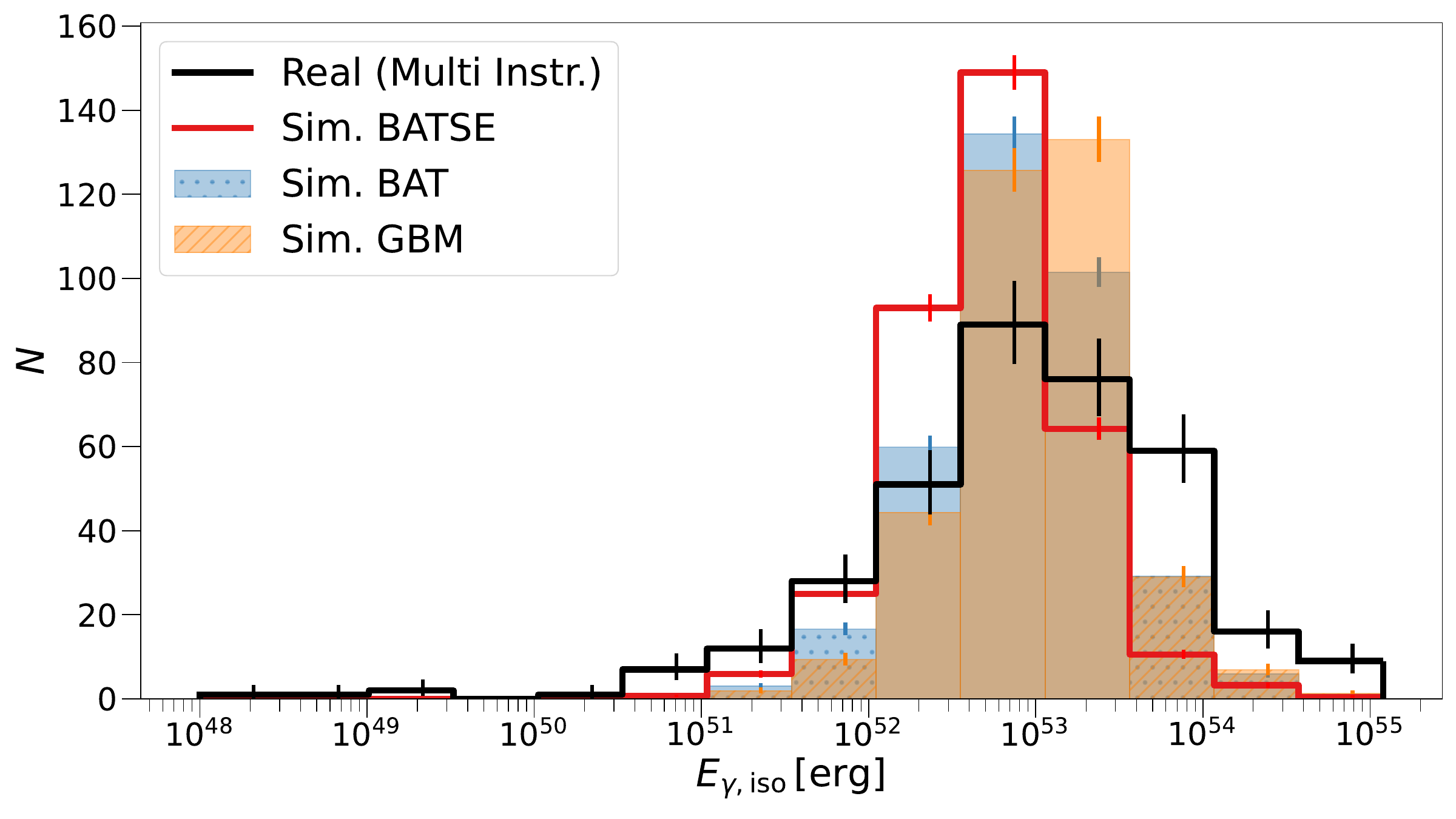}}
    \caption{Bolometric, isotropic-equivalent gamma-ray released energy distributions of the GRB populations simulated for BATSE (red solid line), \textit{Swift}/BAT (shaded blue), and \textit{Fermi}/GBM (shaded orange). The corresponding distribution of real GRBs with known redshift (black solid line) is mostly due to \textit{Swift}, then \textit{Fermi}, Konus/\textit{WIND}, and \textit{BeppoSAX}. The simulated distributions are rescaled to match the total number of observed events.}
    \label{fig:Eiso_distribution}
\end{figure}

\begin{figure}[!ht]
    \resizebox{\hsize}{!}{\includegraphics{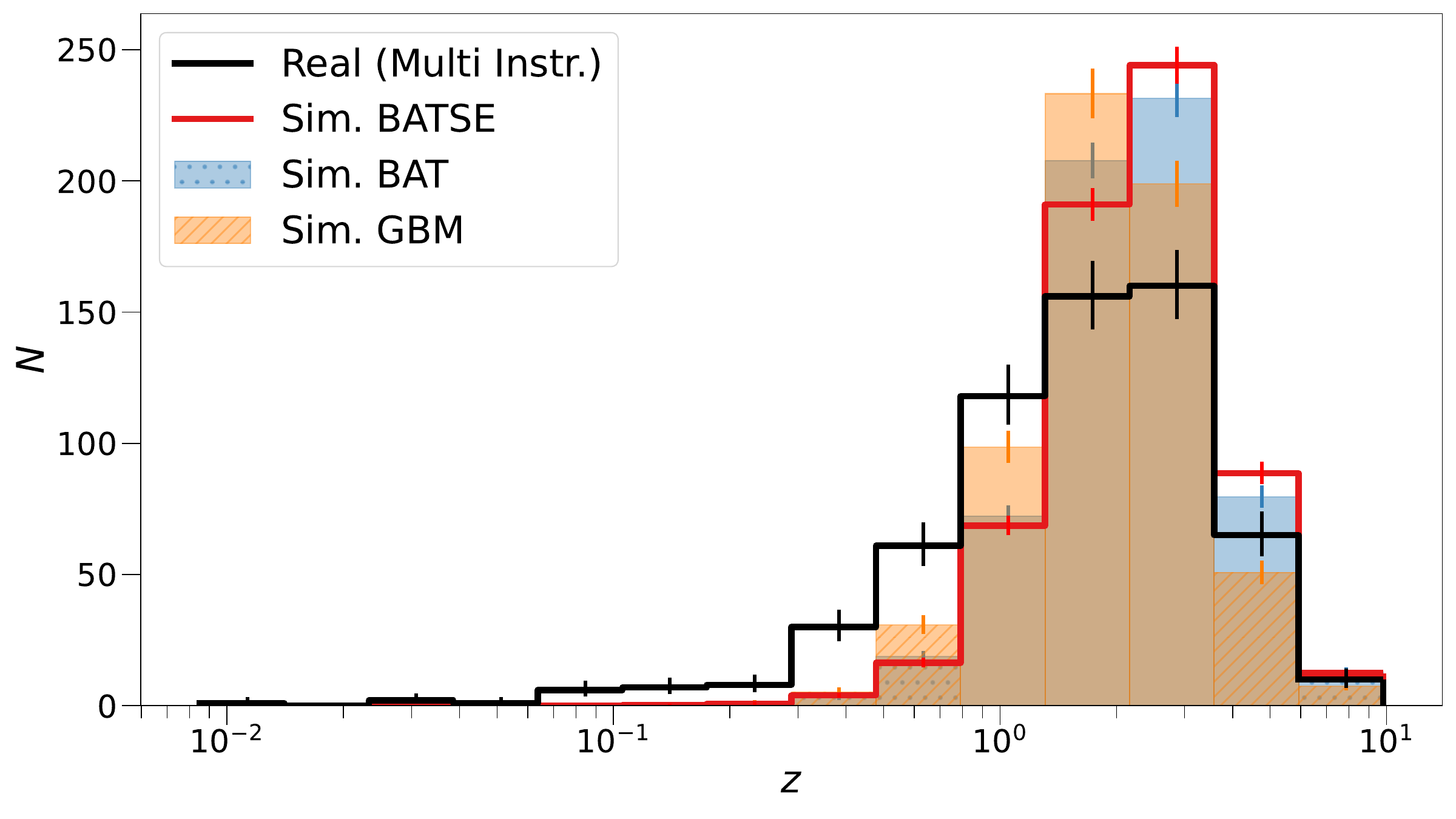}}
    \caption{Simulated redshift distributions inferred from the IS model optimised with BATSE (red line), \textit{Swift} (shaded blue), and \textit{Fermi} samples (shaded orange). The distribution of real GRBs with known $z$ (black solid line) is predominantly due to \textit{Swift}, then \textit{Fermi}/GBM, Konus/\textit{WIND}, and \textit{BeppoSAX}. The simulated distributions are rescaled to match the total number of observed events.}
    \label{fig:z_distribution}
\end{figure}

\begin{figure}[!ht]
    \resizebox{\hsize}{!}{\includegraphics{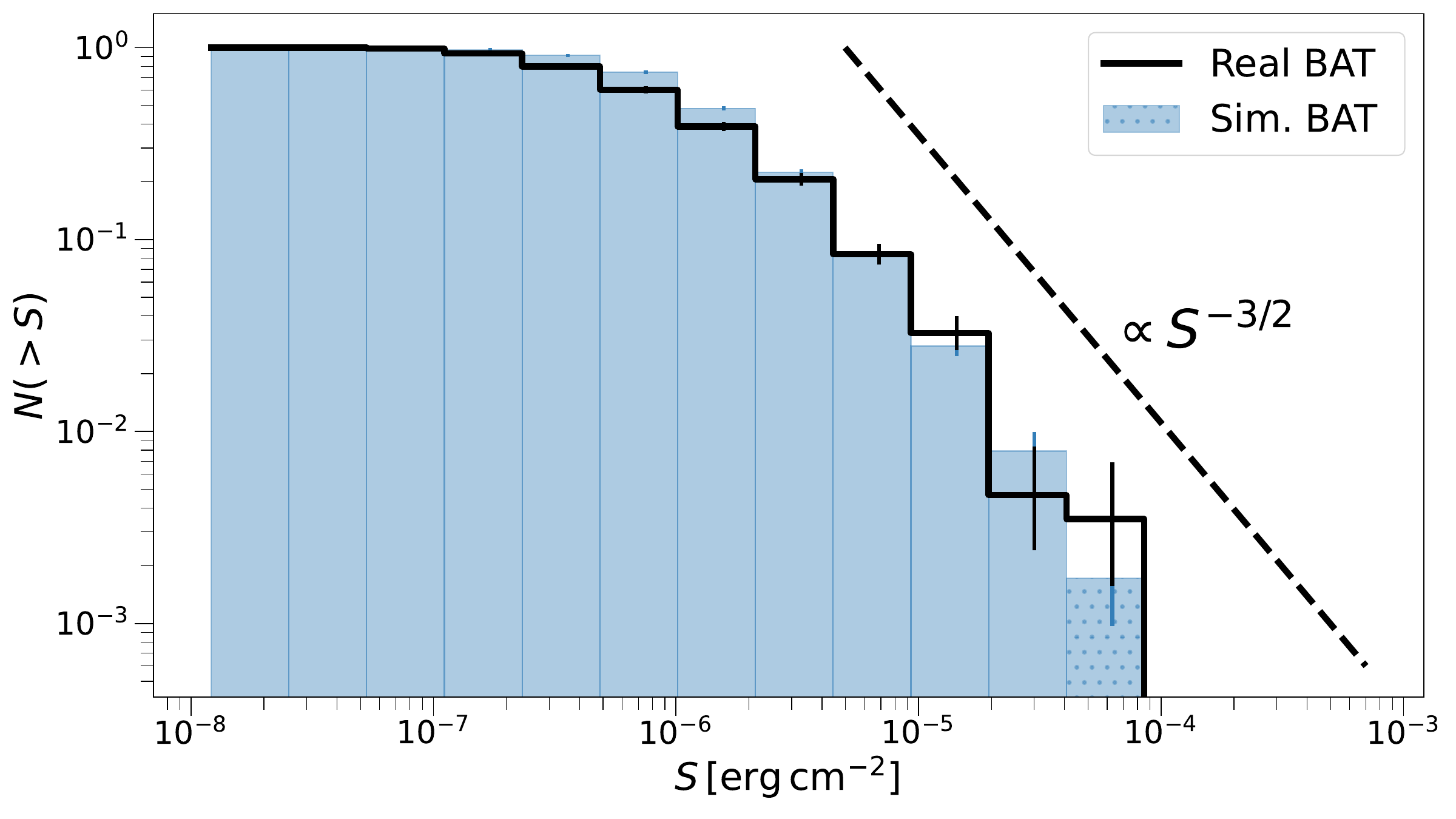}}
    \caption{Simulated (shaded) and real (solid line) 15-150 keV fluence normalised survival distributions of the GRB population seen with \textit{Swift}/BAT. The dashed line indicates the canonical $-3/2$ slope of the $\log N$--$\log S$ GRB distribution for a Euclidean universe.}
    \label{fig:fluence_distribution}
\end{figure}

\begin{figure}[!ht]
    \resizebox{\hsize}{!}{\includegraphics{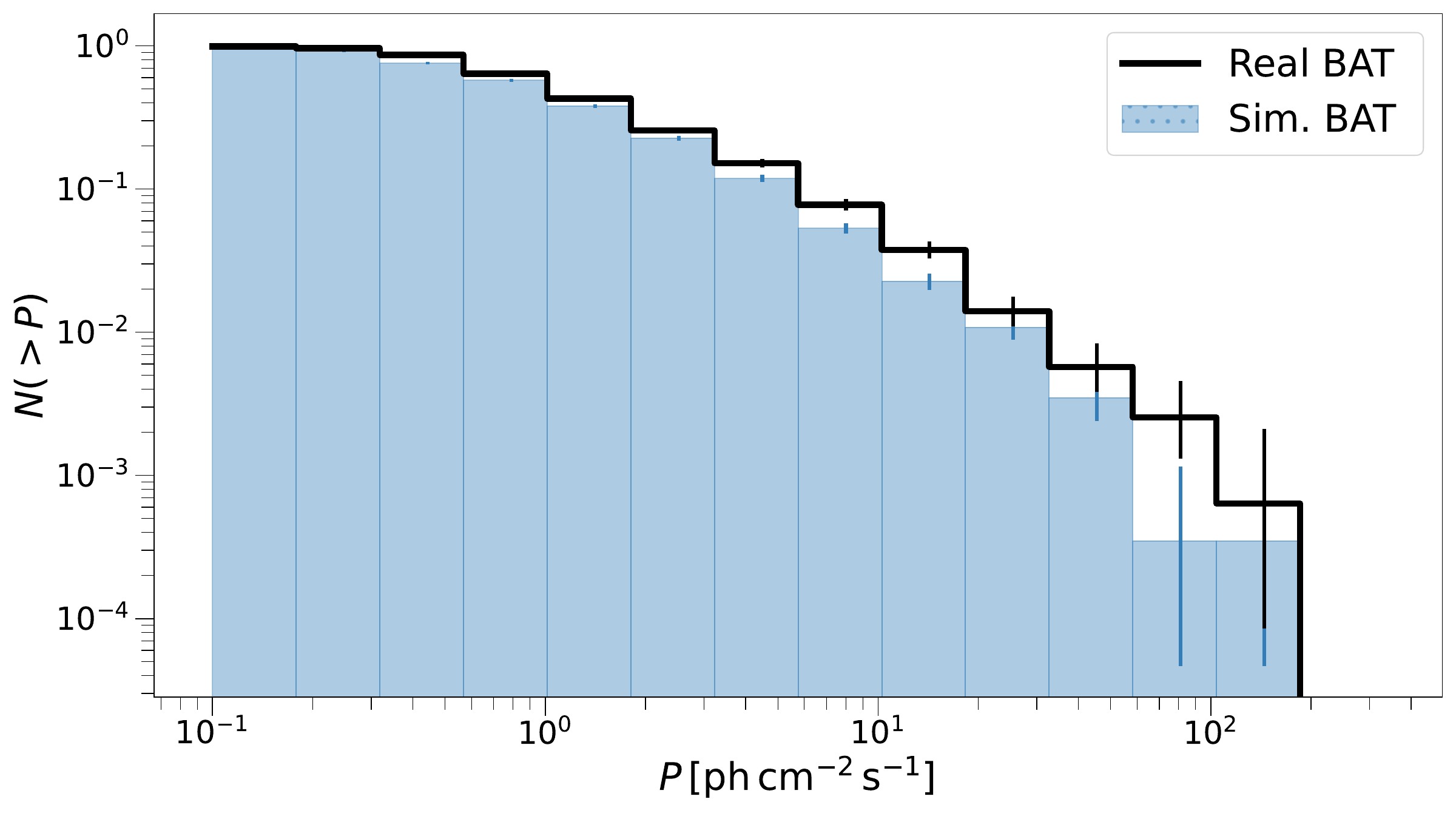}}
    \caption{Simulated (shaded) and real (solid line) 15-150 keV peak photon flux normalised survival distributions of the GRB population seen with \textit{Swift}/BAT. The peak photon flux refers to light curves with a bin time of 1~s.}
    \label{fig:ph_flux_distribution}
\end{figure}

Figures~\ref{fig:Liso_distribution} and~\ref{fig:Eiso_distribution} show the comparisons between simulated and observed distributions of $L_{\gamma,\mathrm{iso}}$ and of $E_{\gamma,\mathrm{iso}}$. All $L_{\gamma,\mathrm{iso}}$ distributions peak in the range $10^{52}$--$10^{53}$ erg~s$^{-1}$, whereas the $E_{\gamma,\mathrm{iso}}$ distributions are centred around $10^{53}$ erg. This overall agreement indicates that the model successfully reproduces the characteristic luminosity and energy ranges observed in real GRBs.
Concerning $E_{\gamma,{\rm iso}}$, the real distribution tends to have more low-energy GRBs ($E_{\gamma,{\rm iso}}\lesssim 10^{51}$~erg~s$^{-1}$).
The redshift distributions are shown in Figure~\ref{fig:z_distribution}. The simulated samples display a broad maximum around $z \sim 2$--3, consistent with the observed distribution of GRBs with known $z$. The relative scarcity of events at $z \lesssim 1$ in the simulations is attributable to the adopted detection criteria; in particular, the observed low-$z$ population includes low-luminosity GRBs (\textit{ll}-GRBs), which preferentially triggered the imaging criteria of \textit{Swift}/BAT.

Figures~\ref{fig:fluence_distribution} and~\ref{fig:ph_flux_distribution} display the predicted cumulative fluence and peak photon flux distributions of the \textit{Swift}/BAT sample compared with the observed ones\footnote{\url{https://swift.gsfc.nasa.gov/archive/grb_table/}} in the corresponding 15-150 keV passband. For reference, the black solid line indicates the canonical $-3/2$ slope expected for the $\log N$--$\log S$ distribution in a Euclidean universe. The simulated distributions closely follow the observed behaviour especially in the bright tail. Overall, these comparisons demonstrate that the model produces simulated populations consistent with the main observational properties of long GRBs.

\subsection{Luminosity function}
\label{sec:lumf}

\begin{figure}[!ht]
    \resizebox{\hsize}{!}{\includegraphics{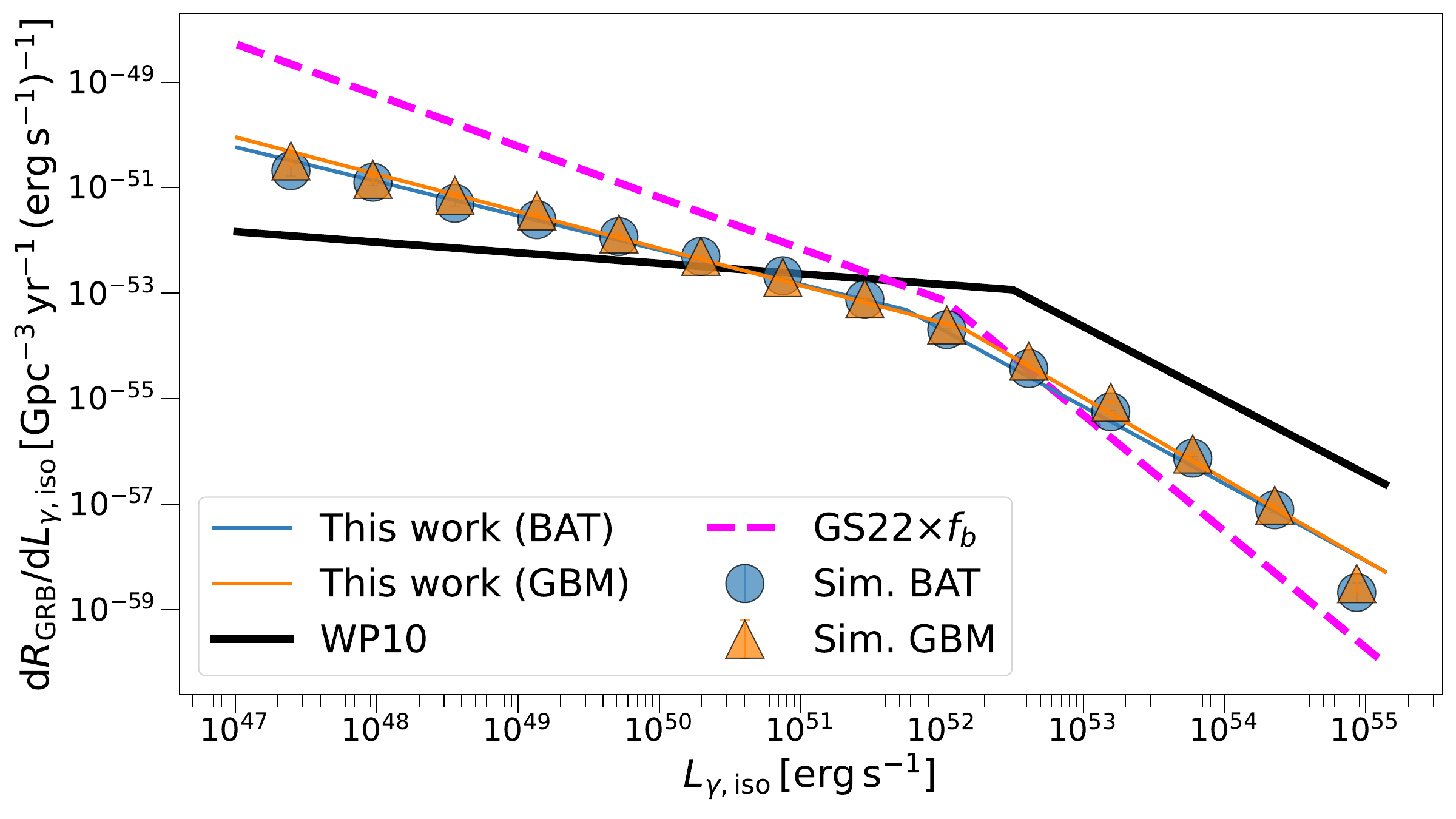}}
    \caption{Local isotropic-equivalent luminosity function obtained from simulated populations based on the optimised IS model calibrated on \textit{Swift} and \textit{Fermi} data. Also shown are the analogous quantity derived from \citet{WandermanPiran10} (solid black) and that by \citetalias{Ghirlanda22} down-scaled by a factor of $f_b^{-1}=100$ (dashed purple), as the original one was corrected for collimation.}
    \label{fig:luminosity_function}
\end{figure}

Figure~\ref{fig:luminosity_function} illustrates the luminosity function $dR_{\rm grb}/dL_{\gamma,{\rm iso}}$, that is, the GRB volumetric rate as a function of $L_{\gamma,{\rm iso}}$, evaluated at $z=0$ under the assumption of no redshift evolution, for the \textit{Swift}/BAT and \textit{Fermi}/GBM samples simulated with the corresponding optimised IS models.
Concerning the \textit{Swift} sample, we counted 505 GRBs detected by BAT with a 15-150-keV peak photon flux $\ge 2.6$~ph~cm$^{-2}$~s$^{-1}$\footnote{This threshold is well above the BAT sensitivity of $\sim 0.3$~ph~cm$^{-2}$~s$^{-1}$ \citep{Lien16}.} over a time span of $21.1$~yr. Given an average duty cycle of $0.75$ and a field of view of $1.4$~sr, the full-sky rate is $286$ GRBs~yr$^{-1}$. The fraction of simulated \textit{Swift}/BAT GRBs with peak photon flux in excess of $2.6$~ph~cm$^{-2}$~s$^{-1}$ and $L_{\gamma,{\rm iso}}>10^{47}$~erg~s$^{-1}$ amounts to $\sim$16\% of the total. Consequently, our IS model, optimised on BAT data, predicts $\sim$1770~GRBs~yr$^{-1}$ with $L_{\gamma,{\rm iso}}>10^{47}$~erg~s$^{-1}$. By demanding that the integral of Eq.~\eqref{eq:Rz} matches this predicted rate, we determined $R_{\rm grb}(0) = 0.13$~Gpc$^{-3}$~yr$^{-1}$, in agreement with previous estimates \citep{Salvaterra12} and somewhat lower than others \citep{Howell14,Sun15,PalmerioDaigne21}. 
For the \textit{Fermi} sample, adopting a peak photon flux threshold of $5$~ph~cm$^{-2}$~s$^{-1}$ in the 10--1000~keV band, together with an average duty cycle of 0.55 and a field of view of 8.8~sr, our model predicts a local rate of $0.15$~Gpc$^{-3}$~yr$^{-1}$.
The un-normalised luminosity function, given by the differential distribution of the simulated $L_{\gamma,{\rm iso}}$ values, was then fitted with a broken power-law:
\begin{equation}
    \dfrac{dN}{dL_{\gamma, \rm iso}} \propto 
    \begin{cases}
        L_{\gamma, \rm iso}^{-a_1} & \text{$L_{\gamma, \rm iso} \le L_{\gamma, \rm iso, b}$}\,,\\
        L_{\gamma, \rm iso, b}^{a_2 - a_1} L_{\gamma, \rm iso}^{-a_2} & \text{$L_{\gamma, \rm iso} > L_{\gamma, \rm iso, b}$}\,,
    \end{cases}
    \label{eq:bpl}
\end{equation}
where $L_{\gamma, \rm iso, b}$ is the break luminosity, and $a_1$ and $a_2$ the power-law indices at low- and high-luminosities, respectively.
The best-fitting parameter values are reported in Table~\ref{tab:luminosity_function}. We obtained the local luminosity function by normalising the simulated distribution and the corresponding best-fit broken power-law, by demanding that the integral matches $R_{\rm grb}(0)$.

In Figure~\ref{fig:luminosity_function} our luminosity function is compared with other two examples: \citet{WandermanPiran10} considered $L_{\gamma,{\rm iso}}$ and found $R_{\rm grb}(0)=1.3_{-0.7}^{+0.6}$~Gpc$^{-3}$~yr$^{-1}$, whereas \citetalias{Ghirlanda22} considered $L_{\gamma}$, that is corrected for collimation. In the latter case, we divided it by a factor of $f_b^{-1}=100$, where $f_b = (1-\cos{\theta_j})$ is the collimation-correction factor, assuming a typical $\theta_j$ of a few degrees. Our luminosity function is broadly consistent with these estimates, exhibiting a similar break luminosity, while the inferred slopes lie between the two reference models.

\begin{table}[!ht]
    \centering
    \setlength{\tabcolsep}{6pt}
    \renewcommand{\arraystretch}{1.4}
    \caption{Best-fitting parameters of the broken power-law model (Eq.~\eqref{eq:bpl}) describing the luminosity function of the simulated $L_{\gamma,\rm iso}$ for the \textit{Swift}/BAT and \textit{Fermi}/GBM samples.}
    \begin{tabular}{c|c|c} \hline\hline
         Parameter                    & \textit{Swift}/BAT      & \textit{Fermi}/GBM \\\hline
         $\log L_{\gamma, \rm iso,b}$ & $51.75_{-0.05}^{+0.07}$ & $52.14_{-0.06}^{+0.05}$ \\
         $a_1$                        & $0.65_{-0.02}^{+0.02}$  & $0.70_{-0.01}^{+0.01}$ \\
         $a_2$                        & $1.46_{-0.02}^{+0.03}$  & $1.55_{-0.03}^{+0.03}$ \\\hline\hline
    \end{tabular}
    \tablefoot{Quoted uncertainties correspond to 90\% credible intervals derived from the posterior distributions obtained through MCMC simulations.}
    \label{tab:luminosity_function}
\end{table}

\section{Discussion}
\label{sec:discussion} 

Several observed properties originally motivated the formulation of the IS model, most notably the lack of temporal pulse broadening within multi-peaked bursts \citep{Fenimore99,RamirezRuiz00}. Early IS-model simulations successfully reproduced the observed average power density spectral slope \citep{Beloborodov98,Beloborodov00a,Spada00} and qualitatively captured the diversity of LCs (\citetalias{Kobayashi97}; \citealt{Daigne98,RamirezRuiz00}). However, because exploring the model complex, multi-dimensional parameter space was computationally prohibitive at the time, these early studies relied largely on fiducial values. Today, the availability of extensive catalogues containing thousands of GRBs, hundreds of which have confirmed redshifts, combined with modern high-performance computing and sophisticated machine learning, allows for a rigorous quantitative optimisation of the IS model. By requiring the simultaneous reproduction of multiple observed properties, this work represents the first comprehensive effort to constrain the IS model parameters in this manner.

Despite the relative simplicity of the \citetalias{Kobayashi97} model, which assumes uniform kinetic energy across all shells and total conversion of dissipated energy into gamma-rays, it successfully reproduces the diverse properties and distributions characteristic of observed LCs. Specifically, the predicted fluence and peak flux distributions are consistent with \textit{Swift} observations (Figures~\ref{fig:fluence_distribution} and \ref{fig:ph_flux_distribution}). One notable exception is the poorer agreement in the BATSE peak-number distribution. This may partly reflect differences in instrumental response, energy band, background treatment, or peak-identification efficiency, and deserves a dedicated investigation.

Leveraging the GRB formation rate from \citetalias{Ghirlanda22}, we constrained several intrinsic properties, finding that the predicted distributions for both $E_{\gamma,{\rm iso}}$ and $L_{\gamma,{\rm iso}}$ of the detectable population align with observations. By accounting for the entire predicted population (independent of experimental detection limits), we derived the luminosity function, defined as the volumetric rate as a function of $L_{\gamma,{\rm iso}}$ for $L_{\gamma,{\rm iso}} > 10^{47}$~erg~s$^{-1}$. Assuming no evolution with redshift beyond a normalisation that scales with the GRB formation rate, Figure~\ref{fig:luminosity_function} presents the local luminosity function ($dR_{\rm grb}/dL_{\gamma,{\rm iso}}$ at $z=0$). This result is well-modelled by a broken power-law, similar to that of \citetalias{Ghirlanda22}, though it exhibits a shallower slope at low luminosities. This discrepancy implies a lower fraction of \textit{ll}-GRBs within the framework of the IS model, suggesting that a significant portion of \textit{ll}-GRBs may have a distinct origin \citep{Campana06,Liang07,Virgili09,Nakar12,Nakar15,Dong23,Irwin25}. Alternatively, if \textit{ll}-GRBs are interpreted as standard GRBs viewed off-axis \citep{Granot05b,Guidorzi09,Pescalli15,Salafia16}, our model’s inability to account for their abundance remains consistent, as it implicitly assumes on-axis observers. The broken-power-law form is not imposed explicitly, but emerges from the combined distributions of the shell properties and collision histories selected by the optimisation. Its detailed physical origin cannot be uniquely identified within the present analysis and would require a further study.

Furthermore, we gained insight into intrinsic distributions that, while less accessible through direct observation, provide a detailed characterisation of the central engine activity: the number of shells per GRB and their rest-frame emission times.

\begin{itemize}
    \item Number of shells ($n$): the number of shells per GRB follows a generalised Zipf distribution (Eq.~\ref{eq:zipf}). Unlike other discrete distributions such as the Poisson, which failed to fit the data, the Zipf-Mandelbrot distribution is heavy-tailed. It is ubiquitous in physics and in many other fields (socio-economic and ecological systems), as it describes many rank-size distributions (see \citealt{DeMarzo21} and references therein). It incorporates a shift term ($S$ in Eq.~\ref{eq:zipf}) representing the number of shells below which the probability levels off; in our datasets, $S$ ranges between 1 and 3. The power-law index $\alpha$\footnote{The power-law index $\alpha$ of the probability mass function must not be confused with the index of the rank-size distribution $\gamma = 1/(\alpha -1) = 1$ for $\alpha=2$ \citep{DeMarzo21}.} of the probability mass function (PMF) of Eq.~\eqref{eq:zipf} for all the datasets is very close to $2$, which is the same as the Gutenberg-Richter law that describes the frequency--size distribution of tectonic earthquakes of any given geographical location.\footnote{In the literature, the Gutenberg-Richter law is often described as a survival distribution, $\log{N(\ge M)} = a - b M$, where $M$ is magnitude and $N(\ge M)$ the number of earthquakes with magnitude equal or greater than $M$, with $b=\alpha - 1$. The $b$-value for tectonic earthquakes is around 1 (that is $\alpha=2)$ \citep{Roberts15}.}
    Just as earthquake frequency scales with magnitude, the frequency of GRBs scales with the number of shells $n$. This power-law nature may indicate self-organised critical (SOC) behaviour, a concept previously proposed for GRB $\gamma$-ray pulses and X-ray flares \citep{WangDai13,Lyu20,Li23,Wei23,Maccary24}.
    
    \item Shell emission times: for any given GRB, the shell emission times follow a negative exponential distribution with an e-folding time $\tau$. This distribution is the hallmark of a simple stochastic process where each shell has a constant, independent probability per unit time of being ejected. The values of $\tau$ are sampled from a log-normal distribution (see Table~\ref{tab:optimised_parameters} for mean and standard deviation). Such a process has been suggested to explain the power-law nature of collective waiting times in $\gamma$-ray pulses and X-ray flares \citep{Guidorzi15b}. If the GRB engine is powered by a hyper-accreting disc, this could imply that the disc is composed of fragments, each with an equal probability per unit time of being accreted and launching a corresponding shell. Theoretically, such disc fragmentation has been identified as a potential driver of central engine activity \citep{Dallosso17,Shahamat21,Lerner26}.
\end{itemize}

Despite these successes, the present implementation involves several simplifying assumptions. Most importantly, the model does not explicitly include the microphysics of relativistic shocks. The relatively low radiative efficiency is a well-known challenge for standard IS models (e.g. \citealt{Kumar99}). Introducing microphysical parameters such as $\epsilon_e$, $\epsilon_B$, and the fraction of accelerated electrons, $\xi$, would allow the radiation spectrum and its characteristic peak energy to be calculated self-consistently. Requiring the spectral peak to lie within the observed range of approximately $0.1$--$1$~MeV could modify the shell properties and other dynamical parameters inferred from the present optimisation, including those reported in Table~\ref{tab:optimised_parameters}. Although the overall dynamical dissipation efficiency of the present model is typically of order 10\%, this quantity should be clearly distinguished from the prompt gamma-ray efficiency, whose self-consistent determination requires modelling how the dissipated energy is transferred to radiating particles and converted into gamma-rays. Such a treatment is beyond the scope of the present work and is left for future investigation.

The model also neglects the relationship between photon arrival times and observed energies arising from latitude-dependent Doppler boosting. Owing to this simplified treatment, spectral evolution is not inherently captured; the dependence of the simulated LCs on the detector energy passband is mediated only through cosmological redshift and time dilation. Nevertheless, the primary objective of this study was to evaluate the ability of the IS model to reproduce the temporal and morphological diversity of GRB LCs, thereby constraining the intrinsic properties of the central engine.

The current framework can be extended to incorporate both shock microphysics and Doppler-related effects. A further development could involve a more realistic collision prescription, or an alternative simplified description of dissipation in an irregular outflow (e.g. \citealt{Kobayashi01}), in which internal energy that is not radiated is converted back into relative bulk motion and dissipated again in subsequent interactions. Such energy recycling could mitigate the reduction in cumulative radiative efficiency when only a fraction of the shock-generated energy is radiated in each collision. By contrast, the present prescription treats collisions as fully inelastic and assumes that all of the resulting internal energy is radiated promptly.

\section{Conclusions}
\label{sec:conc} 

Although the IS model remains one of the most established frameworks for GRB prompt emission, a rigorous optimisation of its parameters to quantitatively reproduce the properties of observed LCs has remained elusive. This work addresses that gap by utilising three independent catalogues comprising several thousand bursts from \textit{Swift}/BAT, \textit{Fermi}/GBM, \textit{CGRO}/BATSE, and optimising the model parameters via a GA. The model successfully accounts for all datasets, yielding parameter values that are broadly consistent across the different catalogues.

Our previous work introduced a stochastic avalanche model (\citealt{Bazzanini24}; \citetalias{Maistrello25}) to reproduce GRB LC distributions through GA optimisation. Based on the runaway pulse mechanism proposed by \citet{SternSvensson96}, that model demonstrated that an avalanche process could underpin multi-peaked GRB structures, though it remained agnostic to the specific physics. The current study advances this effort by replacing the stochastic toy model with a physical IS framework. Combined with the redshift-dependent formation rate from \citetalias{Ghirlanda22}, this physical approach allows the GA to constrain the fundamental parameters governing the central engine's dynamics.

Among its various insights, the model predicts a luminosity function that closely resembles the established broken power-law, with a characteristic break near $10^{52}$~erg~s$^{-1}$. The primary discrepancy lies in the lower predicted volumetric rate of \textit{ll}-GRBs ($L_{\gamma,{\rm iso}} < 10^{50}$~erg~s$^{-1}$). We conclude that the current formulation of the IS model cannot easily account for the inferred abundance of \textit{ll}-GRBs; this suggests a distinct origin for these events, such as shock-breakout radiation or a different geometric configuration (e.g., standard GRBs viewed off-axis).

Future advancements could follow several complementary paths: (i) modelling the energy dependence of the LC by accounting for high-latitude emission, where curvature effects lead to softer spectra due to reduced Doppler boosting; (ii) incorporating the micro-physics of relativistic shocks, assuming synchrotron and inverse Compton radiation, to constrain the micro-physical parameters; (iii) extending this framework to other extensive and upcoming catalogues, including \textit{Insight-HXMT} (Hard X-ray Modulation Telescope; \citealt{Song22_HXMTGRBcatalog}), Konus/\textit{WIND} \citep{KWGRBcat17,KWGRBcat21}, SVOM (Space-based multi-band astronomical Variable Objects Monitor; \citealt{SVOM22}); and (iv) integrating additional metrics that capture the energy-dependent evolution of the LCs.

Finally, the optimised IS model serves as a predictive framework for the populations to be detected by future GRB missions. As a physically grounded tool, the IS model can effectively simulate the LC morphologies of many different GRBs at any redshift. These simulations will be instrumental in testing proposed on-board trigger algorithms. The source code used in this work is publicly available in a GitHub repository\footnote{\url{https://github.com/mmanuele99/internal_shock_grbs}}.

\begin{acknowledgements}
We acknowledge the reviewer for their useful insights and comments.  M.~M. expresses his gratitude to the Astrophysics Research Institute of the Liverpool John Moores University for their warm hospitality during his three-month stay.
\end{acknowledgements}

\bibliographystyle{aa}
\bibliography{alles_grbs}

\begin{appendix}

\section{Multiple-shell interaction}
\label{app:multiple_shell_model}

Here we summarise the numerical procedure used to follow the dynamical evolution of multiple shells.

We consider an outflow composed of $N$ shells, labelled by an index $i = 1, \ldots, N$, according to their emission order from the central engine. Within this framework, the outermost shell corresponds to the first one emitted, while the innermost shell is the most recently emitted. The shell-ejection process is described as a stochastic Poisson process with a time-dependent rate, as introduced in Sect.~\ref{sec:model}, resulting in a higher concentration of shells in the outer regions of the outflow.

Each shell, emitted at a time $t_i$, is characterised by four quantities: the Lorentz factor $\gamma_i$, the mass $m_i = p\gamma_i^{-1}$, the initial radial position $R_i = c\beta_i t_i$, and the width $l_i$. 
The separation from the previously emitted shell $L_i$ is given by $L_i = R_i - R_{i+1} - l_{i+1}$, where the position of the last-emitted shell is set to $R_N = 0$.

The dynamical evolution of the shell train proceeds as follows. Let $j$ denote the index of the $j$th collision, occurring at time $t_j$ and radius $R_{\mathrm{c}}(t_j)$. Given the shell velocities $\beta_i$ and separations $L_i$ at time $t_{j-1}$, the collision time for each adjacent pair satisfying $\beta_{i+1} > \beta_i$ is $\delta t_{i,i+1} = L_i/[c(\beta_{i+1} - \beta_i)]$. Defining $\delta t_j = \min(\delta t_{i,i+1})$, the $j$th collision occurs at $t_j = t_{j-1} + \delta t_j$. Let $s$ and $r$ denote the indices of the slow and rapid shells involved in the collision at time $t_j$. The collision radius is $R_{\mathrm{c}}(t_j) = R_r(t_j) = R_r(t_{j-1}) + c\beta_r\delta t_j$. The radiation produced in this interaction is detected by an observer located at a distance $R_0$ from the central engine at time $t_{\mathrm{obs},j} = [R_0 - R_{\mathrm{c}}(t_j)]/c + t_j$, where $R_0$ is taken to be equal to the total radial extent of the outflow. This time corresponds to the onset of the pulse described by Eq.~\eqref{eq:pulse}.

After the collision, each shell advances from its previous position $R_i(t_{j-1})$ to $R_i(t_j) = R_i(t_{j-1}) + c\beta_i\delta t_j$. All shells retain their Lorentz factor, mass, and width, except for the $s$ and $r$ shells, which merge into a single shell labelled by index $r$, with properties determined as described in Sect.~\ref{sec:model}. The $s$ shell is then removed from the system.

The above procedure is iterated until either all shells have merged into a single shell or the shells become ordered in such a way that their Lorentz factors increase monotonically outward, preventing further collisions.

\section{Spectral parameters}
\label{app:spectral_parameters}

The spectral parameters $\alpha$, $\beta$, and $E_{\mathrm{p,i}}$ of the Band function \citep{Band93}, used to compute the flux correction factor $k_{\mathrm{flux}}$ described in Sect.~\ref{sec:model}, are assigned as follows.

The values of the spectral indices $\alpha$ and $\beta$ were retrieved from the publicly available Fermi GBM Burst Catalog\footnote{\url{https://heasarc.gsfc.nasa.gov/w3browse/fermi/fermigbrst.html}}. We downloaded all GRBs detected between 14 July 2014 and 14 October 2025, and discarded entries that do not satisfy the following constraints \citep{Poolakkil21}:

\begin{itemize}
    \item $A$: positive relative error $< 0.44$ and negative relative error $< 0.83$;
    \item $\alpha$: positive error $< 0.37$ and negative error $< 0.51$;
    \item $\beta$: positive error $< 1.0$ and negative error $< 0.53$;
    \item $E_{\mathrm{p}}$: positive relative error $< 0.35$ and negative relative error $< 0.43$;
\end{itemize}

where $A$ denotes the spectral normalisation. This selection yields a total of 990 sets of spectral parameters, which we refer to as the good sample.

For each simulated GRB, the values of $\alpha$ and $\beta$ are randomly drawn from the good sample. The intrinsic peak energy $E_{\mathrm{p,i}}$ is assigned differently. Simply setting $E_{\mathrm{p,i}} = E_{\mathrm{p}}(1 + z_{\mathrm{GRB}})$, with $E_{\mathrm{p}}$ randomly extracted from the good sample, would neglect the observed $E_{\mathrm{p,i}}$-$L_{\mathrm{iso}}$ correlation \citep{Amati06,Yonetoku04}, where $L_{\mathrm{iso}}$ is the bolometric isotropic-equivalent peak luminosity.

We therefore proceed as follows. Starting from the value of $L_{\mathrm{iso}}$ derived from the theoretical LC, we compute the corresponding $E_{\mathrm{p,i}}$ using the $E_{\mathrm{p,i}}$-$L_{\mathrm{iso}}$ relation reported in \citetalias{Ghirlanda22}, including the intrinsic scatter of the correlation.

The final values of $\alpha$, $\beta$, and $E_{\mathrm{p,i}}$ obtained through this procedure are those assigned to each simulated GRB.

\section{Genetic algorithm and model parameter optimisation}
\label{app:ga}

Genetic algorithms (GAs) are machine-learning algorithms inspired by evolutionary processes, in which a population of candidate solutions evolves over successive generations towards an optimal configuration, as measured by a predefined set of metrics. Within this framework, each candidate solution, or `individual', is represented by a set of eight model parameters (Table~\ref{tab:parameters}), analogous to a genetic chromosome, with each parameter corresponding to a `gene'. The evolution of the population is driven by the relative fitness of individuals, which determines their likelihood of contributing to subsequent generations. Evolutionary processes, such as crossover between parent individuals and random mutations of individual genes, are implemented to promote exploration of the parameter space and prevent premature convergence.

Each individual encodes a specific combination of model parameter values, sampled within the ranges reported in Table~\ref{tab:parameters}. The optimisation proceeds through a fixed sequence of steps, identical to those described in \citetalias{Maistrello25}, which we summarise here for completeness:
\begin{itemize}
    \item the algorithm was run for 30 generations, each consisting of a population of $N_{\mathrm{pop}} = 2000$ individuals.
    \item For each individual, $N_{\mathrm{GRB}} = 2000$ synthetic LCs were generated using the corresponding parameter set. The same selection criteria applied to the real data (Sect.~\ref{ss:sample_selection}) were applied, and simulated LCs failing to meet these requirements were discarded.
    \item The metrics described in Sect.~\ref{ss:metrics} were then computed for the selected simulated LCs and compared with the corresponding metrics derived from the real datasets. The comparison was quantified through an $L_2$ loss function (see \citetalias{Maistrello25} for details). The final loss assigned to each individual was defined as the average of the five metric losses, while the fitness score was taken as the inverse of this quantity.
    \item At each generation, individuals were ranked according to their loss values. New offsprings were produced by combining the genes of the top 15\% fittest individuals. In addition, random mutations were introduced by resampling individual genes from their allowed ranges, with a mutation probability of 4\% per gene.
\end{itemize}

\begin{table}[!ht]
    \setlength{\tabcolsep}{3pt}
    \renewcommand{\arraystretch}{1.4}
    \caption{Parameter space explored in the model optimisation.}
    \label{tab:parameters} 
    \centering 
    \begin{tabular}{c c c c}
        \hline\hline
        Parameter         & Lower bound & Upper bound & Prior \\\hline
        $\alpha$          & $2$         & $4$         & Uniform \\ 
        $S$               & $1$         & $4$         & Uniform \\ 
        $\mu_\tau$        & $0$         & $4$         & Uniform \\
        $\sigma_\tau$     & $0.1$       & $2$         & Uniform \\
        $\gamma_{\min}$   & $10$        & $250$       & Log-uniform \\
        $\gamma_{\max}$   & $350$       & $1000$      & Log-uniform \\
        $p\,(\mathrm{g})$ & $10^{31}$   & $10^{33}$   & Log-uniform \\
        $r$               & $2$         & $20$        & Log-uniform \\\hline\hline 
    \end{tabular}
    \tablefoot{Lower and upper bounds define the parameter ranges explored during the optimisation, while the priors indicate the sampling distributions adopted in the genetic algorithm.}
\end{table}

\section{Examples of light curves}
\label{app:lc_example}

\begin{figure*}[!ht]
    \centering
    \includegraphics[width=\textwidth]{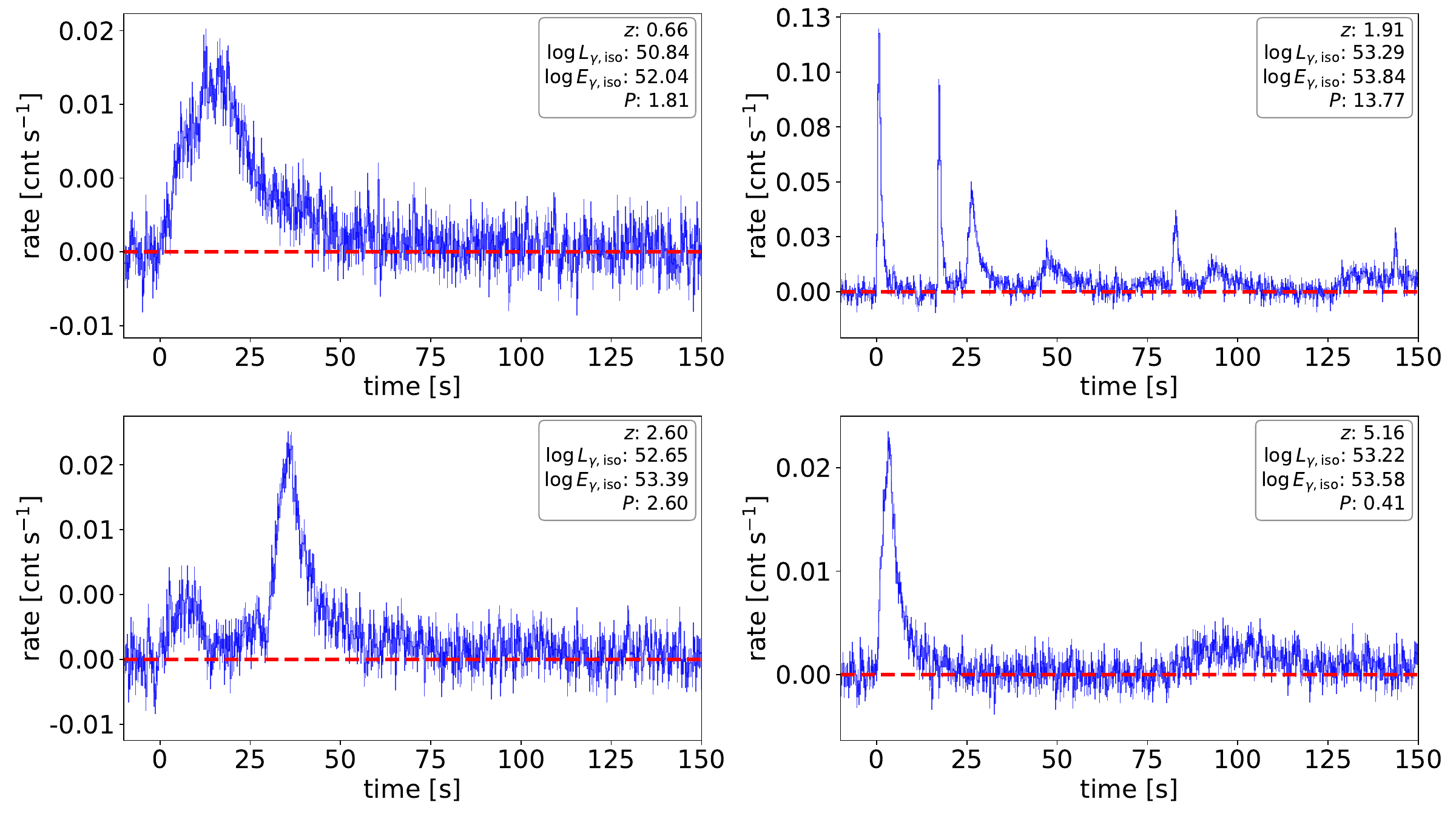}
    \caption{Examples of simulated \textit{Swift}/BAT GRB LCs obtained with the optimised IS model. The bolometric isotropic-equivalent peak luminosity $L_{\gamma, \rm iso}$ and the isotropic-equivalent gamma-ray energy $E_{\gamma, \rm iso}$ are expressed in erg~s$^{-1}$ and erg, respectively, while the peak photon flux $P$ is given in ph~s$^{-1}$~cm$^{-2}$. For each LC, the bin time is 330~ms and corresponds to the shortest detection timescale obtained with {\sc fast-mepsa}, a value that incidentally happens to be common to all four LCs.}
    \label{fig:lc_example}
\end{figure*}

Figure~\ref{fig:lc_example} shows a few examples of simulated \textit{Swift}/BAT GRB LCs obtained with the optimised IS model. Each panel reports the redshift $z$, the bolometric isotropic-equivalent peak luminosity $L_{\gamma, \rm iso}$, the isotropic-equivalent gamma-ray energy $E_{\gamma, \rm iso}$, and the peak photon flux $P$ of the corresponding LC.

\end{appendix}

\end{document}